\documentclass[Times,4pt,aps,pra,onecolumn,showpacs,amsmath,amssymb,floatfix,footinbib,superscriptaddress,times]{revtex4-1}
\usepackage{graphics,graphicx,epsfig,ulem,epstopdf,bm,longtable,url,datetime}
\usepackage[colorlinks=true,linkcolor=red,urlcolor=red,citecolor=red]{hyperref}
\usepackage{amsmath,amssymb,latexsym,float,amsfonts,subfigure,hyperref,color}
\usepackage[ansinew]{inputenc}
\usepackage[usenames,dvipsnames]{pstricks}
\usepackage[mathlines]{lineno}
\def\footnoterule{\kern -1mm \hrule width 5.8cm \kern 2.2mm}%
\usepackage{tikz,xcolor,hyperref}
\definecolor{lime}{HTML}{A6CE39}
\DeclareRobustCommand{\orcidicon}{%
    \begin{tikzpicture}
    \draw[lime, fill=lime] (0,0)
    circle [radius=0.16]
    node[white] {{\fontfamily{qag}\selectfont \tiny ID}};\draw[white, fill=white] (-0.0625,0.095)
    circle [radius=0.007];
    \end{tikzpicture}
    \hspace{-2mm}}
\foreach \x in {A, ..., Z}
{\expandafter\xdef\csname orcid\x\endcsname{\noexpand\href{https://orcid.org/\csname orcidauthor\x\endcsname}{\noexpand\orcidicon}}}

\begin{document}
\title{Left-handedness without absorption in the four-level Y-type atomic medium}

\author{Shun-Cai Zhao\orcidA{}}%
\email[ ]{zscnum1@126.com.}
\affiliation{School of Materials Science and Engineering, Nanchang University, Nanchang 330031, China}
\affiliation{Engineering Research Center for Nanotechnology,Nanchang University, Nanchang 330047,China}
\affiliation{Institute of Modern Physics, Nanchang University, Nanchang 330031,China}

\author{Zheng-Dong Liu}
\email[ ]{lzdgroup@ncu.edu.cn}
\affiliation{School of Materials Science and Engineering, Nanchang University, Nanchang 330031, China}
\affiliation{Engineering Research Center for Nanotechnology,Nanchang University, Nanchang 330047,China}
\affiliation{Institute of Modern Physics, Nanchang University, Nanchang 330031,China}

\author{Qi-Xuan Wu}
\affiliation{College English department, Hainan University, Danzhou 571737, China}

\begin{abstract}
In this paper,three external fields interacting with the
four-level Y-type atomic system described by the density-matrix
approach is investigated .The results show that the left-handedness
with zero absorption are achieved.And the zero absorption property
displays the possibility of manipulation with varying the phase and
the intensity of the coupling field. The zero absorption property
may be used to amplify the evanescent waves that have been lost in
the imaging by traditional lenses.Our scheme proposes an approach to
obtain negative refractive medium with zero absorption and the
possibility to enhance the imaging resolution in realizing
"superlenses".
\end{abstract}

\maketitle
\section{Introduction}
Materials with simultaneous negative electric permittivity and
magnetic permeability have attracted much attention [1-4]. These
materials have extraordinary properties, such as the negative
refraction, inverse Doppler shift and backward-directed Cherenkov
radiation cone, etc.[5].The electric field, the magnetic field and
the wave vector of an electromagnetic wave propagating in such a
material obey the left-hand rule and thus it is called the
left-handed material (LHM) or negative refraction index materials,
which was first introduced by Veselago many years ago[5].In the year
2000, Pendry[6]proposed that a class of ''superlenses'' could be
made by the left-handed materials(LHM). Such lenses may overcome the
traditional limitation in resolution by the wavelength of the used
light and make ''perfect'' images. Since then, the research on such
a superlens and LHM has been booming. Driven by the goals of
realizing a ''perfect lens'', many technologies are used in
demonstrating negative refraction such as composite
materials[1-2,7-17], transmission line simulation[18], photonic
crystal structures [19-24]and photonic resonant materials [25-28].
However, the authors in Ref.[29-30] pointed out that the LHM lenses
can indeed amplify the evanescent waves that have been lost in the
imaging by traditional lenses,the presence of absorption, even a
small amount, plays a crucial rule in recovery of the lost
evanescent waves, thereby controlling the quality of imaging, and
makes LHM lenses less perfect.So, realization of LHM without
absorption is still the key challenge , which is particularly
important in the optical regime [31- 34].

With the realization of LHM without absorption in mind, here we
theoretically put forward a four-level Y-type dense atomic vapor
scheme based on quantum coherence effect to realize left-handedness
with zero absorption.In our scheme,the contribution of
Lorentz-Lorenz local field of the system should be considered.Under
some appropriate conditions,the system shows simultaneously negative
permittivity and permeability,zero absorption.

The paper is organized as follows.In Section 2,we present our model
and its expressions for the electric permittivity ,magnetic
permeability and refractive index.In Section 3, we present numerical
results and their discussion.This is followed by concluding remarks

section{Theoretical model}

\begin{figure}[htp]
\center
\includegraphics[width=0.40\columnwidth]{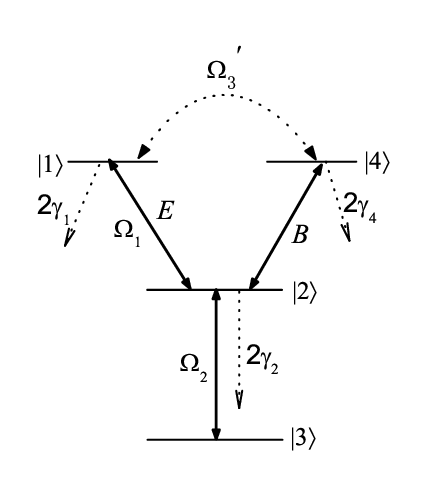}
\caption{The energy level scheme of the four-level Y-type atom driven by the three microwave fields.}
\label{Fig.1}
\end{figure}

Consider a four-level Y-type atomic ensemble interacting with three
optical fields,i.e.the coupling beam, probe light and signal field.
The atomic configuration is schematically shown in figure 1.The
strong resonant coherent field with frequency $\omega_{3}$ and Rabi
frequency $\Omega_{3}$$^{\prime}$ couples states $|1\rangle$ and
$|4\rangle$ (with transition frequency $\omega_{14}$). The weak
probe field with frequency $\omega_{1}$ interacts with the
transition $|1\rangle$ to $|2\rangle$ (transition frequency
$\omega_{12}$),the electric (E) and magnetic (B) components of the
probe field couple state $|2\rangle$ to state $|1\rangle$ by an
electric dipole transition, and to state $|4\rangle$ by a magnetic
dipole transition, respectively.While the weak signal field with
frequency $\omega_{2}$ and Rabi frequency $\Omega_{2}$  drives the
transition $|2\rangle$ to $|3\rangle$( transition frequency
$\omega_{23}$).The parities in this system are $|2\rangle$ even,
$|4\rangle$ even, $|1\rangle$ odd or vice versa. 2 $\gamma_{1}$ and
2 $\gamma_{4}$ are the spontaneous decay rates from the levels
$|1\rangle$ and $|4\rangle$ to the level $|2\rangle$.2 $\gamma_{2}$
corresponds to the decay rate from $|2\rangle$ to $|3\rangle$.The
Hamiltonian under the dipole and rotation-wave approximation is
given by (throughout the paper we assume $\hbar$ = 1) ,

\begin{eqnarray}
\hat{H}= \Delta_{2}|2\rangle \langle2| + (\Delta_{1}
+\Delta_{2})|1\rangle\langle1|+(\Delta_{4}+ \Delta_{2})
|4\rangle\langle4|-(\Omega_{1} |1\rangle\langle2|+ \Omega_{2}
|2\rangle\langle3| \nonumber\\+ \Omega_{3}^{\prime}
|1\rangle\langle4|+H.c.)
\end{eqnarray}
where the detunings of the probe and signal light field are
$\Delta_{1}$=$\omega_{12} -$$\omega_{1}$ and
$\Delta_{2}$=$\omega_{23} -$$\omega_{2}$,respectively. The dynamical
evolution of the system including spontaneous emission is governed
by the equation of motion of the density matrix,
\begin{eqnarray}
\frac{d\rho}{dt}=-\frac{i}{\hbar}[H,\rho]-\sum
\limits_{i,j=1,2,3,4}\gamma_{i,(i\neq3)}\{|j\rangle \langle
j|,\hat{\rho}\} +2\gamma_{1}|2\rangle\langle 1|\rho |1\rangle\langle
2|+2\gamma_{2}|3\rangle\langle 2|\rho |2\rangle\langle 3|
\end{eqnarray}
where [$\hat{A}$,$\hat{B}$] and $\{$$\hat{A}$,$\hat{B}$ $\}$ denote
the commutator and anti-commutator of the operators $\hat{A}$ and
$\hat{B}$, respectively.Here we treat the Rabi frequencies
$\Omega_{1}$ and$\Omega_{2}$ as real
parameters,$\Omega_{3}$$^{\prime}$ as complex parameter:
$\Omega_{3}$$^{\prime}$ = $\Omega_{3}$ $e^{i\Phi_{3}}$, where
$\Phi_{3}$ is the phase of the coupling field. The matrix elements
$\rho_{ij}$ (i, j = 1, 2, 3, 4) obey the differential equations as
follows:
\begin{subequations}
\begin{equation}
\dot{\rho_{11}}=-2\gamma_{1}\rho_{11}-(i \Omega_{1} \rho_{12}+i
\Omega_{3} \rho_{14} e^{-i \Phi_{3}}+H.c.)
\end{equation}
\begin{equation}
\dot{\rho_{33}}=2\gamma_{2} \rho_{22}+(i \Omega_{2} \rho_{23} +
H.c.)
\end{equation}
\begin{equation}
\dot{\rho_{44}}=-2\gamma_{4}\rho_{44}-(i \Omega_{3} \rho_{41}e^{-i
\Phi_{3}}+H.c.)
\end{equation}
\begin{eqnarray}
\dot{\rho_{12}}=-(\gamma_{1}+\gamma_{2}+i\Delta_{1})\rho_{12}-i\Omega_{1}\rho_{11}-i\Omega_{2}\rho_{13}+i\Omega_{1}\rho_{22}+i
\Omega{3} \rho_{42} e^{i \Phi_{3}}
\end{eqnarray}
\begin{equation}
\dot{\rho_{13}}=-(\gamma_{1}+i\Delta_{1}+i\Delta_{2})\rho_{13}-i\Omega_{2}\rho_{12}+i\Omega_{1}\rho_{23}+i
\Omega_{3} \rho_{43}e^{i \Phi_{3}}
\end{equation}
\begin{equation}
\dot{\rho_{14}}=-(\gamma_{1}+\gamma_{4}-i\Delta_{1})\rho_{14}+i\Omega_{1}\rho_{24}-i
\Omega_{3} (\rho_{11}-\rho_{44})e^{i \Phi_{3}}
\end{equation}
\begin{equation}
\dot{\rho_{23}}=-(\gamma_{2}+i\Delta_{2})\rho_{23}+i\Omega_{1}\rho_{13}-i\Omega_{2}\rho_{22}+i\Omega_{2}\rho_{33}
\end{equation}
\begin{equation}
\dot{\rho_{24}}=-(\gamma_{2}+\gamma_{4})\rho_{24}+i\Omega_{1}\rho_{14}+i\Omega_{2}\rho_{34}-i
\Omega_{3} \rho_{21} e^{i \Phi_{3}}
\end{equation}
\begin{equation}
\dot{\rho_{34}}=-(\gamma_{4}-i\Delta_{2})\rho_{34}+i\Omega_{2}\rho_{24}-i
\Omega_{3} \rho_{31} e^{i \Phi_{3}}
\end{equation}
\end{subequations}
where the above density matrix elements obey the conditions:
$\rho_{11}+\rho_{22}+\rho_{33}+\rho_{44}=1$ and
$\rho_{ij}=\rho_{ji}^{\ast}$.

In the following, we will discuss the electric and magnetic
responses of the medium to the probe field.It should be noted that
here the atoms are assumed to be nearly stationary (e.g.,at a low
temperature) and hence any Doppler shift is neglected.When
discussing how the detailed properties of the atomic transitions
between the levels are related to the electric and magnetic
susceptibilities, one must make a distinction between macroscopic
fields and the microscopic local fields acting upon the atoms in the
vapor.In a dilute vapor,there is little difference between the
macroscopic fields and the local fields that act on any
atoms(molecules or group of molecules)[35].But in dense media with
closely packed atoms(molecules),the polarization of neighboring
atoms(molecules) gives rise to an internal field at any given atom
in addition to the average macroscopic field, so that the total
fields at the atom are different from the macroscopic fields[35].In
order to achieve the negative permittivity and permeability,here the
chosen vapor with atomic concentration$N=0.25\times10^{24}m^{-3}$
should be dense,so that one should consider the local field
effect,which results from the dipole-dipole interaction between
neighboring atoms.In what follows we first obtain the atomic
electric and magnetic polarization, and then consider the local
field correction to the electric and magnetic susceptibilities(and
hence to the permittivity and permeability)of the coherent vapor
medium. With the formula of the atomic electric polarizations
$\gamma_{e}=2d_{21}\rho_{12}/\epsilon_{0}E_{p}$,where
$E_{p}=\hbar\Omega_{p}/d_{21}$ one can arrive at
\begin{equation}
\gamma_{e}=\frac{2d_{21}^2\rho_{12}}{\epsilon_{0}\hbar\Omega_{p}}\
\end{equation}

In the similar fashion, by using the formulae of the atomic magnetic
polarizations $\gamma_{m}=2\mu_{0}\mu_{42}\rho_{24}/B_{p}$ [35],and
the relation of between the microscopic local electric and magnetic
fields $E_{p}/B_{p}=c$ we can obtain the explicit expression for the
atomic magnetic polarizability.Where $\mu_{0}$is the permeability of
vacuum,c is the speed of light in vacuum.Then,we have obtained the
microscopic physical quantities $\gamma_{e}$and$\gamma_{m}$ .In
order to achieve a significant magnetic response, the levels
$|1\rangle$ and $|4\rangle$ should be degenerate approximately to
the level $|2\rangle$.However,what we are interested in is the
macroscopic physical quantities such as the electric and magnetic
susceptibilities which are the electric permittivity and magnetic
permeability.The electric and magnetic Clausius-Mossotti relations
can reveal the connection between the macroscopic and microscopic
quantities. According to the Clausius-Mossotti relation [35],one can
obtain the electric susceptibility of the atomic vapor medium
\begin{equation}
\chi_{e}=N\gamma_{e}\cdot{{{{(1-\frac{N\gamma_{e}}{3})}}}}^{-1}
\end{equation}
The relative electric permittivity of the atomic medium reads
$\varepsilon_{r}=1+\chi_{e}$.In the meanwhile,the magnetic
Clausius-Mossotti [36]
\begin{equation}
\gamma_{m}=\frac{1}{N}(\frac{\mu_{r}-1}{\frac{2}{3}+\frac{\mu_{r}}{3}})
\end{equation}
shows the connection between the macroscopic magnetic permeability
$\mu_{r}$ and the microscopic magnetic polarizations $\gamma_{m}$.It
follows that the relative magnetic permeability of the atomic vapor
medium is
\begin{equation}
\mu_{r}=\frac{1+\frac{2}{3}N\gamma_{m}}{1-\frac{1}{3}N\gamma_{m}}
\end{equation}

Substituting the expressions of $\varepsilon_{r}$ and $\mu_{r}$ into
$n=-\sqrt{\varepsilon_{r}\mu_{r}}$[5],we can get the refractive
index of the LHM.In the above,we obtained the expressions for the
electric permittivity ,magnetic permeability and refractive index of
the four-level atomic medium.In the section that follows,we will get
solutions to the density-matrix equations(3)under the stead-state
condition.

\section{Results and discussion}

We consider the atomic vapor medium with $N=0.25\times10^{24}m^{-3}$
to examine the case of dense media where the Lorentz-Lorenz local
field corrections play a significant role.As pointed out in the
following numerical example, the magnitude of the applied coherent
optical field's Rabi frequency is $(10^{6}/s)$,which is larger than
the radiative decay constants.Thus, the effects of dephasing, and
collisional broadening can be neglected,and the coherence effect can
be maintained even increased in the atomic vapor under
consideration. With the steady solution of the matrix equations(3)we
can obtain the coherent terms $\rho_{12}$ and $\rho_{24}$.And with
the expressions for the atomic electric and magnetic
polarizability(4)-(7),we will present a numerical example to show
that the strong electric and magnetic responses can truly arise in
the four-level coherent vapor medium under certain appropriate
conditions. The strong electric and magnetic responses can even lead
to simultaneously negative permittivity and permeability at a
certain wide frequency bands of the probe light.The real parts of
the relative permittivity and permeability,refractive index are
plotted in Figure 2 and Figure 3. We examine the left-handedness and
the absorption behaviors of the four-level Y-type atomic system.

\begin{figure}[htp]
\center
\includegraphics[width=0.40\columnwidth]{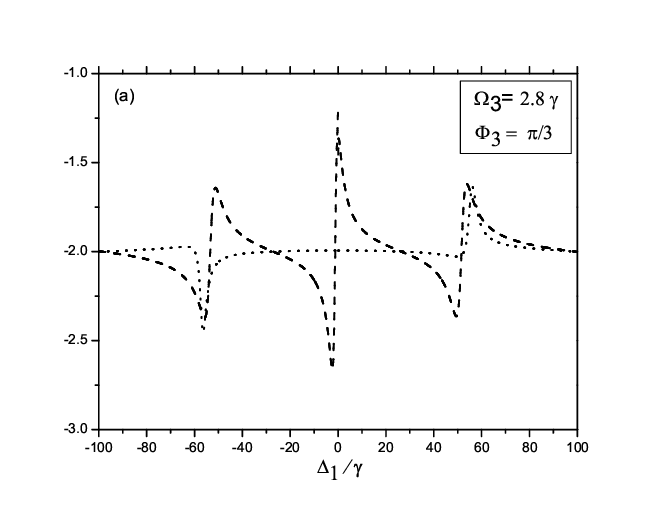}
\includegraphics[width=0.40\columnwidth]{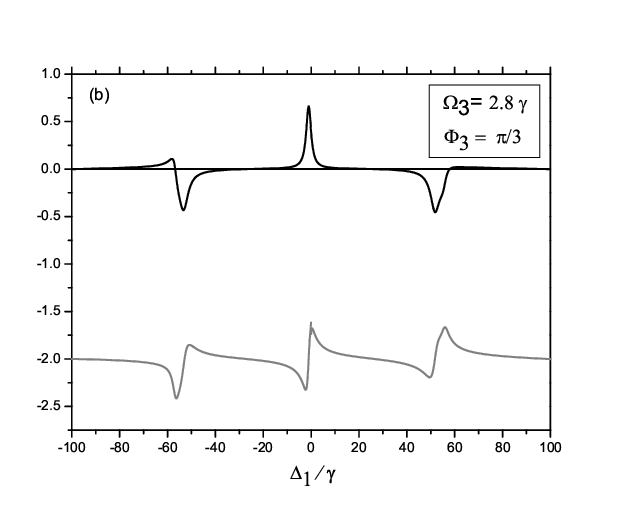}
\caption{(a)Real parts of the permittivity(dash curve)and
permeability(dot curve),(b)real(solid curve)and imaginary(gray
curve) parts of the refractive index as functions of
$\Delta_{1}$/$\gamma$.}
\label{Fig.2}
\end{figure}

In Fig.2,we plot real parts of the permittivity(dash curve) and
permeability(dot curve)[In Fig.2(a)], the refractive index versus
the probe detuning $\Delta_{1}$/$\gamma$ [In Fig.2(b)]with the
parameter values $\Omega_{3}=2.8 \gamma$ and $\Phi_{3}=\pi/3$. The
other parameters are scaled by $\gamma$: $\gamma_{1}$
=$\gamma_{4}$=0.001 $\gamma$, $\gamma_{2}$ =0.005
$\gamma$,$\Omega_{1}$=0.8$\gamma$, $\Omega_{2}$=0.12$\gamma$,
$\Delta_{2}$=0.001 $\gamma$, $\gamma$=10$^{6}$. It can be seen from
Fig.2(a) that the relative dielectric permittivity $\varepsilon_{r}$
has a negative real part in the probe frequency detuning range
[-100$\gamma$;100$\gamma$]. The real part of $\mu_{r}$ is also
negative in these frequency detuning range.The simultaneously
negative permittivity and permeability are exhibited in the
four-level coherent atomic vapor.In Fig.2(b),the gray and solid
curves correspond to the real and imaginary parts of refractive
index.The gray curve shows the real part of refractive index has
negative value in the same probe frequency detuning range as in
Fig.2(a). These mean the coherent atomic vapor becomes an isotropic
left-handed medium with negative permittivity, permeability and
refractive index. The absorption behaviors of the four-level atomic
system are depicted by the imaginary part of refractive index. From
the profile of the solid curve,we can see an absorption peak
approximately at the probe laser resonant location.And the peak
value is about 0.65. On both sides of the absorption peak, there are
two intervals of the probe frequency detuning without absorption and
gain.And the intercals are [-39$\gamma$;-9$\gamma$] and [8$\gamma$;
36$\gamma$]. Beside the two zero absorption intervals appear two
weak gain vales.And the gain ranges are [-74$\gamma$;-39$\gamma$]
and [36$\gamma$;59$\gamma$].The maximum values of the two gain vales

\begin{figure}[htp]
\center
\includegraphics[width=0.40\columnwidth]{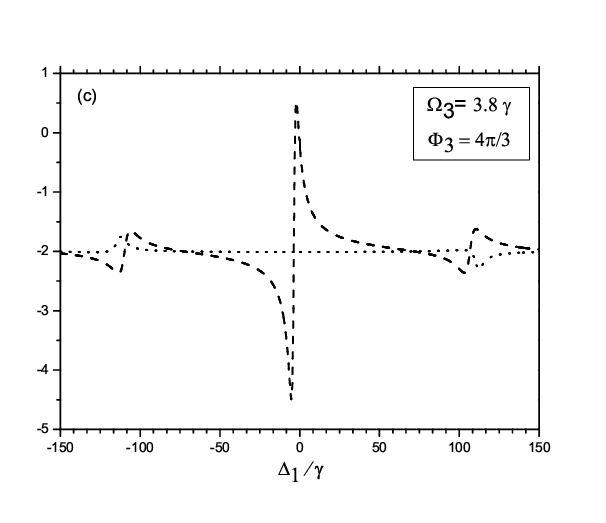}
\includegraphics[width=0.40\columnwidth]{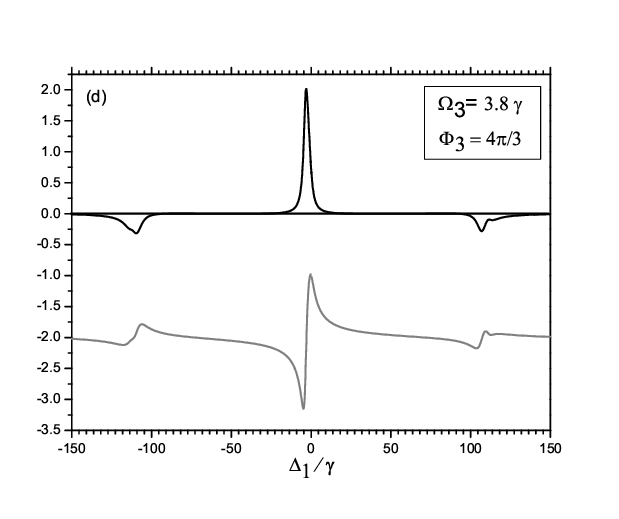}
\caption{(c) Real parts of the permittivity(dash curve)and
permeability(dot curve),(d) real(solid curve)and imaginary(gray
curve)parts of the refractive index as functions of
$\Delta_{1}$/$\gamma$ with other parameters are the same as those in
Figure 2.}
\label{Fig.3}
\end{figure}

In Fig.3,the Rabi frequency and phase of the coupling field  are set
as $\Omega_{3}=3.8 \gamma$ and $\Phi_{3}=4\pi/3$.From Fig.3(c), we
can see the relative dielectric permittivity $\varepsilon_{r}$
exhibits positive real part in the probe frequency detuning range
[-2.5$\gamma$;0].And negative value appears in
[-150$\gamma$;-2.5$\gamma$]and[0;150$\gamma$].The real part of
$\mu_{r}$ is negative in [-150$\gamma$;150$\gamma$].So in the ranges
of [-150$\gamma$;-2.5$\gamma$]and[0;150$\gamma$],the atomic system
displays left-handedness. In Fig.3(d),the profile of the solid curve
shows an absorption peak in the interval of [-17$\gamma$
;12$\gamma$].And the peak value is 2. The zero absorption ranges are
[-98$\gamma$;-17$\gamma$] and [12$\gamma$;96$\gamma$]. Compared
these with Fig.2, the absorption peak value and the zero absorption
intervals are both enlarged in Fig.3. In the ranges of
[-138$\gamma$;-98$\gamma$] and [96$\gamma$;128$\gamma$],two weak
gain vales still appear. The maximum values of the two gain vales
are about -0.3. In other words,increasing the intensity and varying
the phase of the coupling beam can enhance the absorption peak value
at resonant position, extend the zero absorption ranges and reduce
the maximum values the gain values. From the result of Fig.2 and
Fig.3,we both obtained a zero absorption LHM.What's more important
is that we can manipulate the zero absorption property of the LHM
due to the variation of the intensity and phase of the couple
field.The zero absorption property may be used to amplify the
evanescent waves that have been lost in the imaging by traditional
lenses.Our scheme proposes an approach to realize LHM without
absorption and a possibility to enhance the quality of imaging.

\section{Conclusion}

In conclusion,we investigated three external fields interacting with
the four-level Y-type atomic system.Considering the Rabi frequency
of the coupling field as complex parameter, the atomic system can
display left-handedness with simultaneous negative permittivity and
negative permeability,zero absorption property.Varying the intensity
and phase of the couple field can change the absorption properties
at the probe light resonant position, extend the zero absorption
ranges and reduce the gain values beside the zero absorption
intervals.The variational properties manifest the controllable
possibility .The zero absorption property may be used to amplify the
evanescent waves that have been lost in the imaging by traditional
lenses. Our scheme proposes a possibility  to enhance the imaging
resolution in realizing "superlenses".

\section*{Acknowledgments}

The work is supported by the National Natural Science Foundation of
China ( Grant No.60768001 and No.10464002 ).

\end{document}